%% file: report.tex
\documentclass[12pt,dvips,a4paper]{article}
\usepackage[pdftex]{graphicx}
\include{commands}
\usepackage{amssymb}
\usepackage{amsmath}
\usepackage{setspace}
\usepackage{wasysym} 
\usepackage[numbers]{natbib}

\begin{document}

\renewcommand{\title}{Lattice modelling of corrosion induced cracking and bond in reinforced concrete}

\begin{center} \textbf{\title} \end{center}

\begin{center} Peter Grassl$^*$ and Trevor Davies\\
  University of Glasgow\\
  Glasgow, United Kingdom\\
$^*$ corresponding author. Email: peter.grassl@glasgow.ac.uk
\end{center}

\begin{center}
Submitted to Cement and Concrete Composites, 21 March 2011\\
\end{center}

\section*{Abstract}
A lattice approach is used to describe the mechanical interaction of a corroding reinforcement bar, the surrounding concrete and the interface between steel reinforcement and concrete.
The cross-section of the ribbed reinforcement bar is taken to be circular, assuming that the interaction of the ribs of the deformed reinforcement bar and the surrounding concrete can be captured by a cap-plasticity interface model.
The expansive corrosion process is represented by an Eigenstrain in the lattice elements forming the interface between concrete and reinforcement.
Several pull-out tests with varying degree of corrosion are analysed. The numerical results are compared with experiments reported in the literature.
The influence of the properties of concrete are studied. 
The proposed lattice approach offers insight into corrosion induced cracking and its influence on bond strength.

\section{Introduction}
Corrosion of reinforcement involves the transformation of steel into rust, which is an expansive process \cite{bro97}.
If the expansion is restrained, it induces radial pressure in the confining material. 
For reinforced concrete, the radial pressure and accompanying transverse tensile stress may cause cracking \cite{AndAloMol93}. 
Cracking is not desirable because it reduces the anchorage capacity of the reinforcement \cite{AlsKalBasRas90, LeeNogTom02, FanLunPlo06}.
Most of the anchorage capacity of deformed reinforcement is provided by ribs on the surface of the bar, which resist the slip between concrete and reinforcement by transferring inclined radial forces into the concrete~\cite{Tep79}.
The capacity of the concrete to resist these forces can be significantly reduced by corrosion-induced cracking.
Consequently, there is a considerable interest in developing models of corrosion-induced cracking, which can quantify the influence of cracking on the bond capacity of reinforced concrete.

\begin{figure}
\begin{center}
\includegraphics[width=5cm]{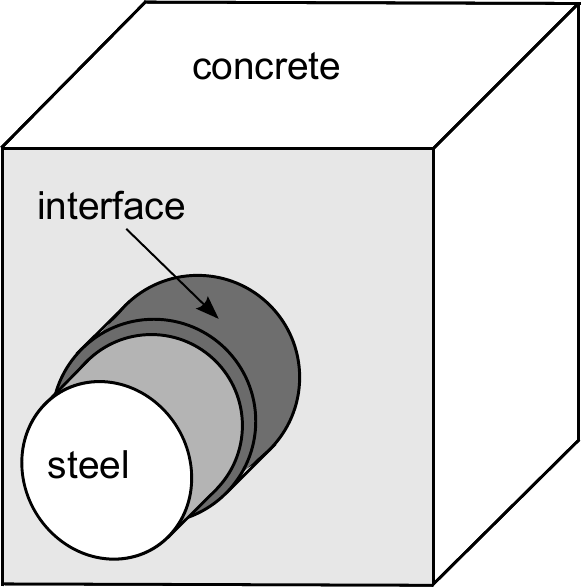}\\
\end{center}
\caption{Three-dimensional modelling of reinforced concrete: Concrete, reinforcement and bond between concrete and reinforcement are considered as individual phases.}
\label{fig:3DBond}
\end{figure}

The mechanics of corrosion-induced cracking and its influence on bond properties are complex. 
For example, at the microscale, the compaction and penetration of rust into pores and micro-cracks takes place \cite{OugBerFraFoc06, WonZhaKar10}. 
Modelling at this scale is challenging, since the microstructure of concrete is complex. 
Therefore, most research is concentrated either at the mesoscale, where the interaction of the ribs of the reinforcement and the heterogeneous concrete are described explicitly, or the macro-scale, where these interactions are described by an interface constitutive model. 
At the macro-scale, the formation of rust is modelled by an expansion of the interface, which results in  macroscopic cracking and subsequent spalling.
Many of the models at the macro-scale proposed in the literature include the effect of corrosion-induced cracking on bond by reducing the bond strength of the interface between concrete and steel \cite{LeeNogTom02}.
In these models, the relationship between the amount of rust and the reduction of bond strength is determined empirically.
Thus, these models are of limited validity for the prediction of the influence of corrosion on bond.
Other models describe the expansion of the rust, the radial pressures and the transverse stresses on the concrete explicitly \cite{Lun07}.
These models have the potential to establish an analytical relationship between the expansion of rust, cracking and spalling.
They can be combined with realistic bond models \cite{Lun05}, so that the influence of corrosion-induced cracking on the bond capacity can be predicted.
However, this modelling framework is computationally demanding, since it requires three-dimensional modelling of the mechanical response of the concrete, the bond between reinforcement bar and concrete, and the reinforcement bar itself, as shown in Fig.~\ref{fig:3DBond}.

In view of these difficulties, discrete methods appear to offer a favourable alternative approach, as they are known to be efficient for the modelling of displacement discontinuities at interfaces \cite{Kaw77}. 
Discrete methods can be subdivided into two categories: particle models and lattice models.
In particle models, the arrangement of particles can evolve, so that former neighbours may become separated. Therefore, such models
are suitable for describing processes involving large deformations.
On the other hand, in lattice models the connectivity between nodes is unchanged during analysis, so that contact determination is not required.
Consequently, lattice models are mainly suitable for analyses involving small strains \cite{HerHanRou89,SchMie92b,BolSai98}.
Their potential to model corrosion-induced cracking and its influence on bond is assessed in the present study.

\section{Lattice modelling approach}
Lattice approaches have been used successfully in the past to model the failure of concrete \cite{SchMie92b,BolSai98}.
In these approaches, the lattice elements do not represent the meso-structure of the material, but instead discretise the continuum.
For example, Bolander and his coworkers have accurately reproduced analytical solutions for elasticity and potential flow problems \cite{YipMohBol05,BolBer04}.
Lattice models can also incorporate constitutive models, formulated in terms of tractions and displacement jumps, as commonly used in interface approaches
for concrete fracture \cite{CabLopCar06}. These models yield element size-independent descriptions of crack-openings \cite{GraJir10}. 
The spatial arrangement of the lattice elements and their cross-sectional properties is based on Delaunay and Voronoi tesselations of a set of random points placed in the domain \cite{YipMohBol05}.
The random placement of nodes reduces the influence of the discretisation on the fracture patterns, as observed for other fracture approaches \cite{GraRem07,JirGra08}.
For concrete, we adopt a damage-plasticity constitutive model following the work presented in \cite{GraJir06, GraRem08, Gra09b}.
For the bond between reinforcement and concrete, a non-associated plasticity interface model is proposed, which is conceptually similar to the model developed by Lundgren \cite{Lun05}.
This model consists of a Mohr-Coulomb friction law combined with a compression cap.
For the steel phase, an elastic constitutive model is used.

\begin{figure}[h!]
\begin{center}
\includegraphics[width=5cm]{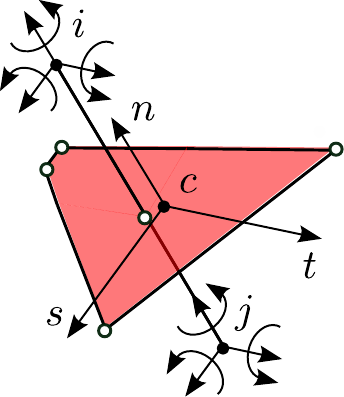}\\
\end{center}
\caption{3D lattice element.}
\label{fig:element}
\end{figure}

For each two-noded lattice element, a local co-ordinate system is introduced (Fig.~\ref{fig:element}): the axis $n$ is aligned with the element axis and axes $s$ and $t$ are aligned with the two principal directions of the cross-section of the lattice element.    
Each node has six degrees of freedom (three translations and three rotations) which are used to determine the displacement jump at the centroid $c$ of the element cross-section in the local coordinate system by rigid body motions as 
\begin{equation}\label{eq:kinematics}
\mathbf{u}_{\rm c} = \mathbf{B} \mathbf{u}_{\rm e}
\end{equation}     
where $\mathbf{u}_{\rm c} = \left\{ u_{\rm cn}, u_{\rm cs}, u_{\rm ct}, \phi_{\rm n}, \psi_{\rm s}, \theta_{\rm t} \right\}^T$ are the displacement and rotation discontinuities at the point $c$ and $\mathbf{u}_{\rm e} = \left\{u_{1}, v_{1}, w_{1}, \phi_{1}, \psi_{1}, \theta_{1}, u_{2}, v_{2}, w_{2}, \phi_{2}, \psi_{2}, \theta_{2} \right\}^T$ are the degrees of freedom at the two nodes.
Furthermore, the matrix $\mathbf{B}$ in Eq.~(\ref{eq:kinematics}) is
\begin{equation}
\mathbf{B} =
\left [ \begin{array}{cccccccccccc}
-1 & 0 & 0 & 0 & -e_{\rm t} & e_{\rm s} & 1 & 0 & 0 & 0 & e_{\rm t} &-e_{\rm s}\\
0 & -1 & 0 & e_{\rm t} & 0 & -h/2 & 0 & 1 & 0 & -e_{\rm t} & 0 & -h/2\\
0 & 0 & -1 & -e_{\rm s} & h/2 & 0 & 0 & 0 & 1 & e_{\rm s} & h/2 & 0\\
0 & 0 & 0 & -\sqrt{\dfrac{I_{\rm p}}{2 A}} & 0 & 0 & 0 & 0 & 0 & \sqrt{\dfrac{I_{\rm p}}{2 A}} & 0 & 0\\
0 & 0 & 0 & 0 & -\sqrt{\dfrac{I_1}{A}} & 0 & 0 & 0  & 0 & 0 & \sqrt{\dfrac{I_1}{A}} & 0\\
0 & 0 & 0 & 0 & 0 & -\sqrt{\dfrac{I_2}{A}} & 0 & 0  & 0 & 0 & 0 & \sqrt{\dfrac{I_2}{A}}
\end{array}
\right ]
\label{eq:Bmatrix}
\end{equation}

In Eq.~(\ref{eq:Bmatrix}), $e_{\rm s}$ and $e_{\rm t}$ are the eccentricities between the midpoint of the lattice element and the centre $c$ in the directions $s$ and $t$, respectively. In addition, $h$ is the length of the element and $A$ is the cross-sectional area. Furthermore, $I_{\rm p}$ is the polar second moment of area and $I_1$ and $I_2$ are the two principal second moments of area of the cross-section.
In the local coordinate system, the stiffness matrix is:
\begin{equation}
\mathbf{K}_{\rm e} = \dfrac{A}{h} \mathbf{B}^{\rm T} \mathbf{D} \mathbf{B}
\end{equation}
where $\mathbf{D}$ is the 6x6 constitutive matrix of the material.
In the following sections, the constitutive models for bond, concrete and reinforcement are described for the displacement discontinuities $\bar{\mathbf{u}}_{\rm c} = \left\{u_{cn}, u_{cs}, u_{ct}\right\}^{T}$.

\subsection{Constitutive model for the bond between concrete and reinforcement} \label{bondModel}

\begin{figure}
\begin{center}
\includegraphics[width=10cm]{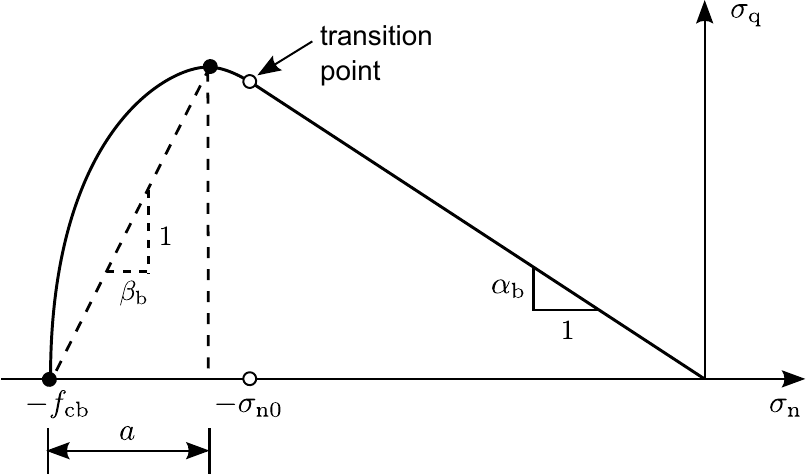}
\end{center}
\caption{Yield surface: Mohr-Coulomb friction law combined with a cap.}
\label{fig:surface}
\end{figure}

The interface response between reinforcement and concrete is characterised by displacement jumps which are related to tractions via an interface constitutive model. 
The three-dimensional displacement jump $\bar{\mathbf{u}}_{\rm c} = \left\{u_{\rm{cn}}, u_{\rm{cs}}, u_{\rm{ct}}\right\}^{T}$ is transformed into strains $\boldsymbol{\varepsilon} = \left\{\varepsilon_{\rm n}, \varepsilon_{\rm s}, \varepsilon_{\rm t} \right\}^T$ by means of the interface thickness $h$ as
\begin{equation}\label{eq:smear}
\boldsymbol{\varepsilon} = \dfrac{\bar{\mathbf{u}}_{\rm c}}{h}
\end{equation}
The thickness of the interface $h$ is chosen to be equal to the length of the lattice element, which for the bond model crosses the interface between
the reinforcement steel and the concrete.
The strain $\boldsymbol{\varepsilon}$ is related to the stress $\boldsymbol{\sigma} = \left(\sigma_{\rm n}, \sigma_{\rm s}, \sigma_{\rm t} \right)^T$ by the elasto-plastic stress-strain relationship
\begin{equation}\label{eq:stressStrain}
\boldsymbol{\sigma} = \mathbf{D}_{\rm e} \left(\boldsymbol{\varepsilon} - \boldsymbol{\varepsilon}_{\rm p} - \boldsymbol{\varepsilon}_{\rm{cor}}\right)
\end{equation}
where $\mathbf{D}_{\rm e}$ is the elastic stiffness, $\boldsymbol{\varepsilon}_{\rm p} = \left(\varepsilon_{\rm pn}, \varepsilon_{\rm ps}, \varepsilon_{\rm pt} \right)^T$ is the plastic strain and $\boldsymbol{\varepsilon}_{\rm{cor}} = \left(\varepsilon_{\rm{cor}},0, 0 \right)^T$ is the Eigenstrain used to describe the expansion associated with the corrosion process (see Sec.~\ref{sec:corrosion}).
The elastic bond stiffness matrix is
\begin{equation}
\mathbf{D}_{\rm e} = \begin{pmatrix} 
E_{\rm b} & 0 & 0\\
0 & \gamma_{\rm b} E_{\rm b} & 0\\
0 & 0 & \gamma_{\rm b} E_{\rm b} \end{pmatrix}
\end{equation}
where $E_{\rm b}$ is the normal stiffness and $\gamma_{\rm b}$ is the ratio of shear and normal stiffnesses.
The yield surface of the plasticity model consists of a Mohr-Coulomb friction law combined with an elliptical cap. 
The shape of the cap surface is adjusted so that a smooth transition between the two surfaces is obtained (Fig.~\ref{fig:surface}). This combination was initially proposed by \cite{SwaSeo00} for a circular cap and further developed by \cite{DolIbr07}.
The yield function $f$ depends on the normal stress $\sigma_{\rm n}$ and the shear stress norm $\sigma_{\rm q} = \sqrt{\sigma_{\rm s}^2 + \sigma_{\rm t}^2}$ as
\begin{equation}\label{eq:yieldInter}
f = \left\{ \begin{array}{ll} 
  \sigma_{\rm q} + \alpha_{\rm b} \sigma_{\rm n} &  \mbox{if $\sigma_{\rm n0} \leq \sigma_{\rm n}$} \vspace{0.5cm}\\
  \sigma_{\rm q}^2 + \dfrac{\left(\sigma_{\rm n} + f_{\rm c} -a \right)^2}{\beta_{\rm b}^2} - \dfrac{a^2}{\beta_{\rm b}^2} & \mbox{if $\sigma_{\rm n} \leq \sigma_{\rm n0}$} \vspace{0.5cm}
\end{array} \right.
\end{equation}
\noindent where $\alpha_{\rm b}$ is the friction angle and $f_{\rm c}$ is the compressive strength of concrete.
Furthermore,
\begin{equation} 
a = \dfrac{\beta_{\rm b} \alpha_{\rm b} f_{\rm c}}{\alpha_{\rm b} \beta_{\rm b} + \sqrt{1+\beta_{\rm b}^2\alpha_{\rm b}^2}}
\end{equation}
where $\beta_{\rm b}$ is the ratio of the short and long axes of the cap ellipse (Fig.~\ref{fig:surface}).
At the point where the two parts of the yield surface meet, the normal stress is  
\begin{equation}
\sigma_{\rm n0} = -\dfrac{a}{ \beta_{\rm b} \alpha_{\rm b} \sqrt{1+\beta_{\rm b}^2 \alpha_{\rm b}^2 }}
\end{equation}

The rate of the plastic strains in Eq.~(\ref{eq:stressStrain}) is
\begin{equation}\label{eq:flow}
\dot{\boldsymbol{\varepsilon}}_{\rm p} = \dot{\lambda} \dfrac{\partial g} {\partial \bar{\boldsymbol{\sigma}}}
\end{equation}
where $g$ is the plastic potential and $\lambda$ is the plastic multiplier.
In the present study, $g$ is chosen to be very similar to the yield function $f$. 
The only difference is that $\alpha_{\rm b}$ is replaced by a flow inclination $\psi_{\rm b}$ so that the magnitude of the normal plasticity strain generated during shear loading can be controlled.
Thus, the plastic potential is
\begin{equation}\label{eq:plast}
g = \left\{ \begin{array}{ll} 
  \sigma_{\rm q} + \psi_{\rm b} \sigma_{\rm n} &  \mbox{if $\sigma_{\rm n0g} \leq \sigma_{\rm n}$} \vspace{0.5cm}\\
  \sigma_{\rm q}^2 + \dfrac{\left(\sigma_{\rm n} + f_{\rm c} -a_{\rm g} \right)^2}{\beta_{\rm b}^2} - \dfrac{a_{\rm g}^2}{\beta_{\rm b}^2} & \mbox{if $\sigma_{\rm n} \leq \sigma_{\rm n0g}$} \vspace{0.5cm}
\end{array} \right.
\end{equation}
with
\begin{equation} 
a_{\rm g} = \dfrac{\beta_{\rm b} \psi_{\rm b} f_{\rm c}}{\psi_{\rm b} \beta_{\rm b} + \sqrt{1+\beta_{\rm b}^2\psi_{\rm b}^2}}
\end{equation}
and
\begin{equation}
\sigma_{\rm n0g} = - \dfrac{a_{\rm g}}{\beta_{\rm b} \psi_{\rm b} \sqrt{1+ \beta_{\rm b}^2 \psi_{\rm b}^2}}
\end{equation}

The plasticity model is completed by the loading and unloading conditions:
\begin{equation}\label{eq:loadUn}
f \leq 0 \mbox{,} \hspace{0.5cm} \dot{\lambda} \geq 0 \mbox{,} \hspace{0.5cm} \dot{\lambda} f = 0
\end{equation}
This plasticity bond model is similar to the one developed by Lundgren \cite{Lun05} for bond between concrete and reinforcement, 
but there are several differences. Here, a smooth transition between the cap and the frictional law is introduced, so that a special vertex stress return
algorithm in the transition region is obviated. 
Also, the response is perfectly-plastic. Thus, the calibration is different from the one used for the model proposed by Lundgren.

\subsection{Model for corrosion between concrete and reinforcement}\label{sec:corrosion}

\begin{figure}
\begin{center}
\begin{tabular}{cc}
\includegraphics[width=3.5cm]{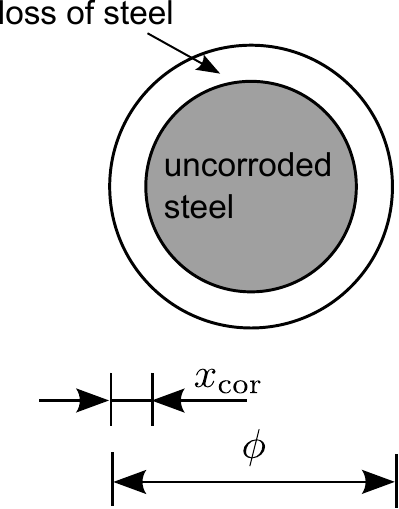} \hspace{1cm}& \includegraphics[width=5.cm]{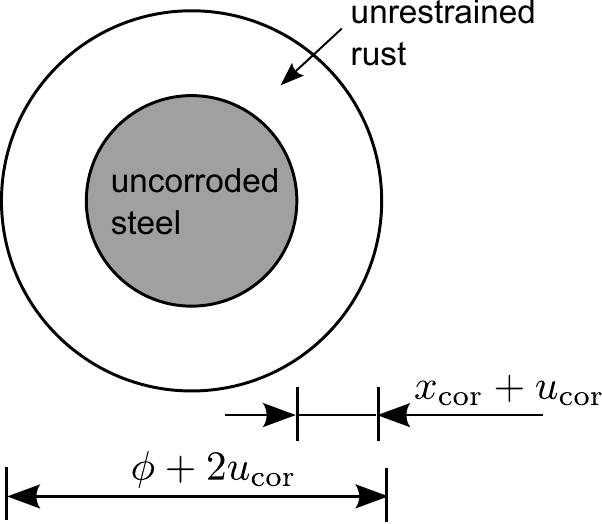}\\
(a) \hspace{1cm} & (b)
\end{tabular}
\end{center}
\caption{Representation of the corrosion process as an expansive layer of rust: a) steel loss, b) unrestrained expansion of rust}
\label{fig:corrosion}
\end{figure}

The effect of corrosion is modelled by an Eigenstrain $\varepsilon_{\rm{cor}}$ in Eq.~(\ref{eq:stressStrain}), which is determined from the free expansion of the corrosion product $u_{\rm{cor}}$ as
\begin{equation}
\varepsilon_{\rm{cor}} = \dfrac{u_{\rm{cor}}}{h}
\end{equation}
where $h$ is the length of the element. For the bond model, this is the length of the element across the concrete-steel interface.
This approach has been shown to give results which are independent of the element length \cite{Gra09}.
The free expansion $u_{\rm{cor}}$ is determined from the corrosion penetration $x_{\rm{cor}}$, which is the thickness of  the layer of steel that is lost during the corrosion process (Fig.~\ref{fig:corrosion}a).
The percentage of steel loss $\rho$ is
\begin{equation} \label{eq:density}
\rho = \dfrac{\Delta V_{\rm s}}{V_{\rm s}}\times 100 = \dfrac{\pi \phi^2/4 - \pi \left(\phi-2 x_{\rm{cor}}\right)^2/4}{\pi \phi^2/4} \times 100
\end{equation}
where $V_{\rm s}$ is the cross-sectional area of the uncorroded reinforcement bar, $\Delta V_{\rm s}$ is the cross-sectional area that is lost during the corrosion process and $\phi$ is the diameter of the reinforcement bar.
Solving Eq.~(\ref{eq:density}) for $x_{\rm{cor}}$ gives
\begin{equation}
x_{\rm{cor}} = \phi/2 \left(1-\sqrt{1 - \rho/100}\right)
\end{equation} 
The transformation of the steel layer $x_{\rm{cor}}$ into rust is assumed to be accompanied by a cross-sectional area expansion
\begin{equation}\label{eq:expansion}
\Delta V_{\rm{cor}} = \lambda_{\rm{cor}} \Delta V_{\rm{s}}
\end{equation}
assuming an unrestrained expansion (Fig.~\ref{fig:corrosion}b). Here $\lambda_{\rm{cor}}$ is a model parameter.
The cross-sectional area of the rust can be written as
\begin{equation}\label{eq:corrosionArea}
\Delta V_{\rm{cor}} = \pi \left(\phi + 2u_{\rm{cor}}\right)^2/4 - \pi \left(\phi - 2x_{\rm cor}\right)^2/4
\end{equation}
where $u_{\rm{cor}}$ is the expansion of the unrestrained corrosion layer as shown in Fig.~\ref{fig:corrosion}b.
Equating Eqs.~(\ref{eq:expansion})~and~(\ref{eq:corrosionArea}), substituting for $\Delta V_{\rm s}$ from Eq.~(\ref{eq:density}) and solving for $u_{\rm cor}$ gives
\begin{equation}
u_{\rm{cor}} = \sqrt{\phi^2/4 + \left(\phi x_{\rm{cor}} - x_{\rm{cor}}^2\right)\left(\lambda_{\rm{cor}}-1\right)} - \phi/2
\end{equation}
The bond model, which considers corrosion,  has seven parameters, namely $E_{\rm b}$, $\gamma_{\rm b}$, $f_{\rm c}$, $\alpha_{\rm b}$, $\beta_{\rm b}$, $\psi_{\rm b}$ and $\lambda_{\rm cor}$. The calibration of these parameters is discussed later.

\subsection{Constitutive model for concrete} \label{sec:concreteModel}
The constitutive model for concrete is based on a damage-plasticity framework.
The strains, which are determined from the displacement jumps as discussed earlier, are related to the nominal stress $\boldsymbol{\sigma} = \left(\sigma_{\rm n}, \sigma_{\rm s}, \sigma_{\rm t} \right)^T$ as 
\begin{equation}\label{eq:totStress}
\boldsymbol{\sigma} = \left(1-\omega\right) \mathbf{D}_{\rm e} \left(\boldsymbol{\varepsilon}-\boldsymbol{\varepsilon}_{\rm p}\right) = \left(1-\omega\right) \bar{\boldsymbol{\sigma}}
\end{equation}
where $\omega$ is the damage variable, $\mathbf{D}_{\rm e}$ is the elastic stiffness, $\boldsymbol{\varepsilon}_{\rm p} = \left(\varepsilon_{\rm pn}, \varepsilon_{\rm ps}, \varepsilon_{\rm pt} \right)^T$ is the plastic strain and $\bar{\boldsymbol{\sigma}}$ is the effective stress.
The elastic stiffness has the same format as the one for the bond model. The two parameters denoted $E_{\rm c}$ and $\gamma_{\rm c}$ control the Young's modulus and Poisson's ratio of the material. 
The plasticity model for concrete is based on the effective stress and is thus independent of damage. 

\begin{figure}
\begin{center}
\includegraphics[width=12cm]{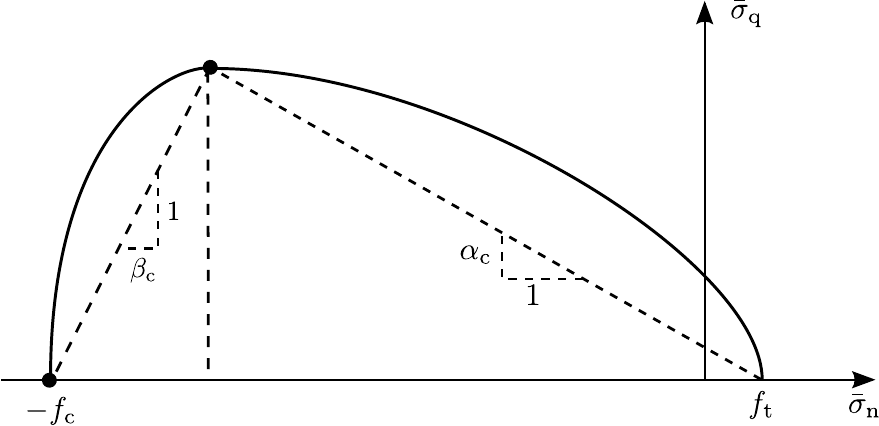}
\end{center}
\caption{Yield surface: The yield surface is controlled by the tensile strength $f_{\rm t}$, the compressive strength $f_{\rm c}$ and two parameters $\alpha_{\rm c}$ and $\beta_{\rm c}$.}
\label{fig:yieldSurfaceConcrete}
\end{figure}

Again, the plasticity part is described by the yield function (Eq.~\ref{eq:yieldInter}), the flow rule (Eq.~\ref{eq:plast}) and the loading unloading conditions (Eq.~\ref{eq:loadUn}) as described for the bond plasticity model.
The yield function, shown in Fig.~\ref{fig:yieldSurfaceConcrete} for $f = 0$, which is a function of two stress variables $\bar{\sigma}_{\rm n}$ and $\bar{\sigma}_{\rm q} = \sqrt{\bar{\sigma}_{\rm s}^2 + \bar{\sigma}_{\rm t}^2}$, is defined as 
\begin{equation}\label{eq:yieldCon}
f = \left\{ \begin{array}{ll} \alpha_{\rm c}^2\bar{\sigma}_{\rm n}^2 + 2 \dfrac{\alpha_{\rm c}^2\left(f_{\rm c} - \alpha_{\rm c} \beta_{\rm c} f_{\rm t}\right)}{1+\alpha_{\rm c} \beta_{\rm c}} \bar{\sigma}_{\rm n} + \bar{\sigma}_{\rm q}^2 - \dfrac{2 \alpha_{\rm c}^2 f_{\rm c}f_{\rm t} + \alpha_{\rm c}^2 \left(1-\alpha_{\rm c} \beta_{\rm c} \right) f_{\rm t}^2}{1+\alpha_{\rm c} \beta_{\rm c}} & \hspace{0.5cm} \mbox{if $\bar{\sigma}_{\rm n} \geq \dfrac{f_{\rm c} - \alpha_{\rm c} \beta_{\rm c} f_{\rm t}}{1+\alpha_{\rm c} \beta_{\rm c}} $} \vspace{0.5cm}\\
\dfrac{\bar{\sigma}_{\rm n}^2}{\beta_{\rm c}^2} + 2 \dfrac{f_{\rm c} - \alpha_{\rm c} \beta_{\rm c} f_{\rm t}}{\beta_{\rm c}^2 \left(1+\alpha_{\rm c} \beta_{\rm c}\right)} \bar{\sigma}_{\rm n} + \bar{\sigma}_{\rm q}^2 + \dfrac{\left(1-\alpha_{\rm c} \beta_{\rm c}\right) f_{\rm c}^2 -2 \alpha_{\rm c} \beta_{\rm c} f_{\rm c} f_{\rm t} }{\beta_{\rm c}^2 \left(1+\alpha_{\rm c} \beta_{\rm c} \right)}  & \hspace{0.5cm} \mbox{if $\bar{\sigma}_{\rm n} < \dfrac{f_{\rm c} - \alpha_{\rm c} \beta_{\rm c} f_{\rm t}}{1+\alpha_{\rm c} \beta_{\rm c}} $} \end{array} \right.
\end{equation}
where $f_{\rm t}$ is the tensile strength, $f_{\rm c}$ is the compressive strength, and $\alpha_{\rm c}$ and $\beta_{\rm c}$ are the inclinations shown in Fig.~\ref{fig:yieldSurfaceConcrete}. 
The plastic potential $g$ is almost identical to the yield surface, except that the inclination $\alpha_{\rm c}$ is replaced by $\psi_{\rm c}$.

The damage parameter in Eq.~(\ref{eq:totStress}) is determined by means of the damage history variable
\begin{equation} \label{eq:damageHistory}
\kappa_{\rm d} = \left \langle \varepsilon_{\rm pn} \right \rangle
\end{equation}
where $\langle . \rangle$ denotes the McAuley brackets (positive part of operator). 
The relation of this history variable to the damage parameter is derived from the response in pure tension for which $\sigma_{\rm n} > 0$ and $\sigma_{\rm q} = 0$.
For this stress state, the crack opening is defined as 
\begin{equation}
w_{\rm c} = h \left( \varepsilon_{\rm pn} + \omega \left(\varepsilon_{\rm n} - \varepsilon_{\rm pn} \right)\right)
\end{equation} 
where $h$ is the length of the lattice element.
The damage law is based on an exponential stress-crack opening relationship in the post-peak regime for the normal stress component. 
It has the form 
\begin{equation} \label{eq:softLaw}
\sigma_{\rm n} = f_{\rm t} \exp \left(-\dfrac{w_{\rm c}}{w_{\rm f}}\right)
\end{equation}
where $w_{\rm f}$ is a model parameter, which is related to the local fracture energy $G_{\rm F}$ of the concrete lattice elements as $w_{\rm f} = G_{\rm F}/f_{\rm t}$.
The normal stress component is also prescribed in Eq.~(\ref{eq:totStress}) as
\begin{equation} \label{eq:totStressNormal}
\sigma_{\rm n} = \left(1-\omega\right) E_{\rm c} \left(\varepsilon_{\rm n} - \varepsilon_{\rm pn}\right)
\end{equation}
In Eq.~(\ref{eq:totStressNormal}), the elastic strain $\varepsilon_{\rm n} - \varepsilon_{\rm pn}$ in the post-peak regime can be expressed as $f_{\rm t}/E_{\rm c}$, since the plasticity model is perfect plastic.
Equating Eqs.~(\ref{eq:softLaw})~and~(\ref{eq:totStressNormal}), and using Eq.~(\ref{eq:damageHistory}), a nonlinear equation for the damage parameter $\omega$ is obtained as
\begin{equation}
\left(1-\omega\right) = \exp \left( - \dfrac{h \left(\kappa_{\rm d} + \omega f_{\rm t}/E_{\rm c}\right)}{w_{\rm f}} \right)
\end{equation}
 From this equation, the damage parameter $\omega$ is determined iteratively by means of a Newton iteration. 
The eight parameters for the constitutive model of concrete are $E_{\rm c}$, $\gamma_{\rm c}$, $f_{\rm t}$, $f_{\rm c}$, $\alpha_{\rm c}$, $\beta_{\rm c}$, $\psi_{\rm c}$ and $w_{\rm f}$.

\subsection{Constitutive model for the reinforcement}
The material response of the reinforcement is modeled by the linear elastic stress-strain relationship
\begin{equation}
\boldsymbol{\sigma} = \mathbf{D}_{\rm e} \boldsymbol{\varepsilon}
\end{equation}
where the two parameters of the elastic stiffness are denoted as $E_{\rm s}$ and $\gamma_{\rm s}$.

\section{Comparison with experimental data}

\begin{figure}[ht!]
\begin{center}
\begin{tabular}{cc}
\includegraphics[width=9.cm]{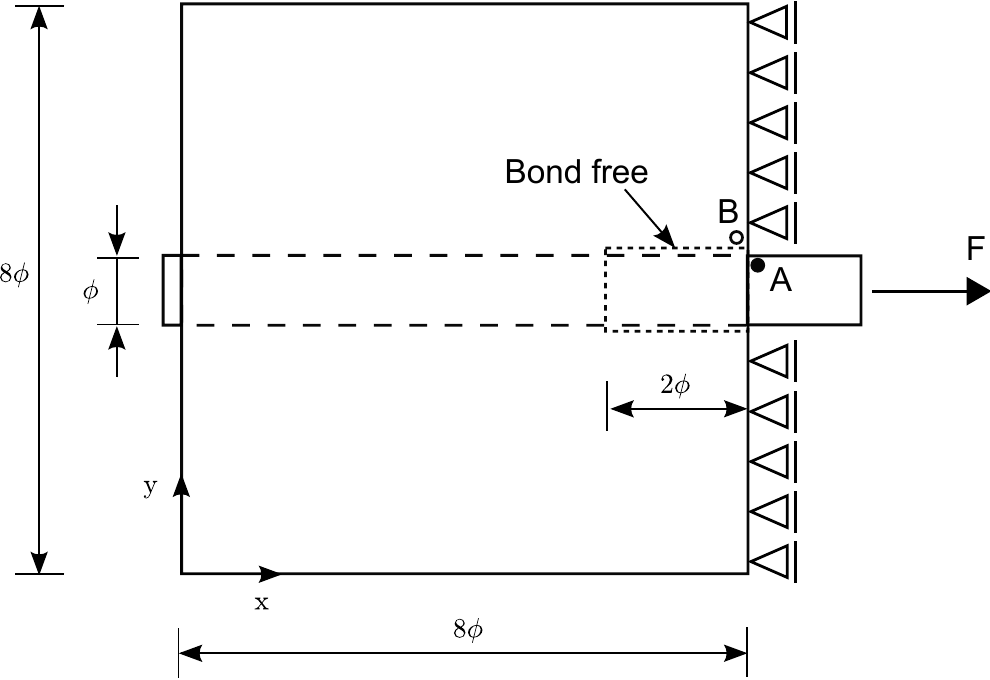} & \includegraphics[width=6.cm]{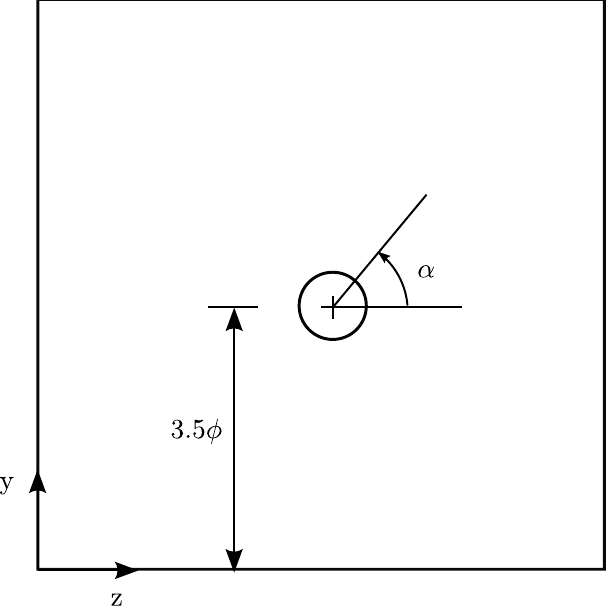}\\
(a) & (b)
\end{tabular}
\end{center}
\caption{Geometry and loading set-up for the corrosion pull-out tests reported by \protect \cite{LeeNogTom02}. The reinforcement bar (diameter $\diameter = 13$~mm)  is placed eccentrically.}
\label{fig:geometry}
\end{figure}

The lattice approach described in the previous section is used to model the experiments reported in \cite{LeeNogTom02}.
The geometry and loading setup of the experiments are shown in Fig.~\ref{fig:geometry}.
Reinforcement bars ($\diameter = 13$~mm) embedded in concrete cubes were initially subjected to corrosion and subsequently pulled out.
In the analyses, it was assumed that the cubes were restrained on one face by friction-less supports, which were modelled by roller boundary conditions in the numerical analyses. All other faces are assumed to be traction-free.

\begin{figure}
\begin{center}
\includegraphics[width=7cm]{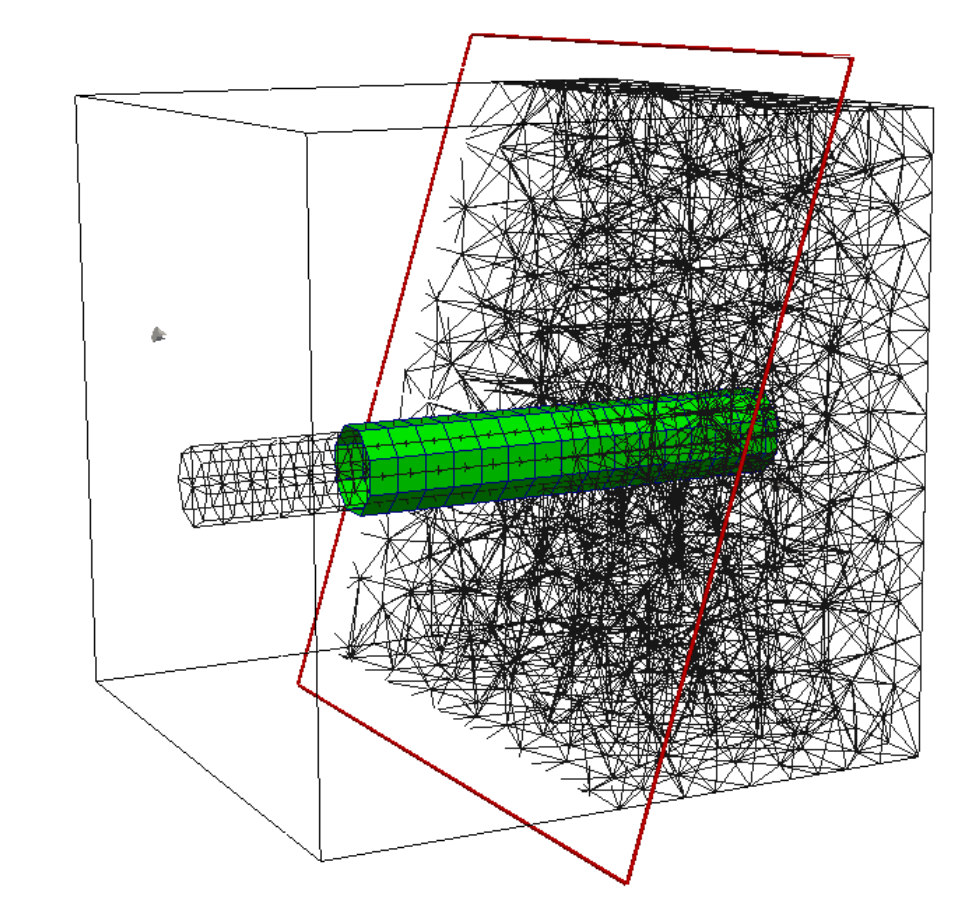}
\end{center}
\caption{A part of the mesh for the lattice analysis.}
\label{fig:mesh}
\end{figure}

The concrete used in the experiments is characterised by a Young's modulus of $22.6$~GPa, Poisson's ratio of $0.17$, a tensile strength of $2.7$~MPa
 and a compressive strength of $24.7$~MPa. The Young's modulus of the reinforcement is $183$~GPa.
The response of the concrete, reinforcement and bond between concrete and reinforcement is modelled using the lattice shown in Fig.~\ref{fig:mesh}, which consists of 9165 lattice elements.
 For the reinforcement and the interface between the reinforcement and the concrete the mesh is structured, but for the concrete itself the lattice is randomly orientated.
The rectangles in Fig.~\ref{fig:mesh} are the cross-sections of the lattice elements which represent the interface between the reinforcement bar and the concrete. For these elements, the element axis is normal to the interface. Therefore, the Eigenstrain normal component $\varepsilon_{\rm cor}$ represents the strain in the radial direction.
Furthermore, the cap-plasticity bond model for these elements is used to idealise the interaction of the ribs of the deformed bar and the surrounding concrete, which are not modelled explicitly.

Three tests were analysed.
In the first test, the reinforcement was pulled out without initial corrosion.
In the other two tests, corrosion levels of $\rho = 3.2$ and $16.8$~\% were reached before the pullout. 
In all three analyses, the load $F$ was controlled by imposing an end slip, which is defined as the horizontal displacement of node $A$ as shown in Fig.~\ref{fig:geometry}a.

The proposed model for corrosion-induced cracking explicitly represents the three phases, (a) concrete, (b) reinforcement and (c) bond between concrete and reinforcement. 
Three constitutive models are required for this purpose, and inevitably this demands specification of several parameters; 17 in total.
To determine these parameters,  the following calibration strategy was adopted.
Firstly, the two parameters for the reinforcement were calibrated based on a cube specimen so that the stiffness of the reinforcement was matched.
This gives $E_{\rm s} = 345$~GPa and $\gamma_{\rm s} =0.065$.
The same specimen was used to calibrate the elastic parameters of the concrete model, which were determined to be $E_{\rm c} =36.6$~GPa 
and $\gamma_{\rm c} =0.175$. 
In the next step, a uniaxial tension and compression test was carried out to determine the remaining parameters of the concrete model.
 Two uniaxial tests are not sufficient to determine 6 model parameters uniquely. Therefore, the parameters $\alpha_{\rm c} = 0.5$, $\beta_{\rm c} = 0.5$ and $\psi_{\rm c} = 0.25$ are assumed to be default values. The tensile and compressive strength were set to $f_{\rm t} = 2.2$~MPa and $f_{\rm c} = 40$~MPa.
Furthermore, the parameter $w_{\rm f} = 0.045$~mm was determined based on the assumption of a local fracture energy of $G_{\rm F} = 100$~N/m. 

For the bond model, the elastic parameters were chosen as $E_{\rm b} = 2E_{\rm s} E_{\rm c}/(E_{\rm c}+E_{\rm s})$ and $\gamma_{\rm b} = 0.175$. Here, $E_{\rm b}$ is the harmonic mean of $E_{\rm s}$ and $E_{\rm b}$ assuming that half of the element crossing the interface is steel and the other half is concrete.
The other parameters of the bond model were calibrated in two steps. First, the frictional parameter $\alpha_{\rm b} = 0.24$ and dilation parameter $\psi_{\rm b} = 0.05$ were calibrated so that the peak value of the experimental stress displacement curve for $\rho=0$ was obtained. The corrosion related parameter $\lambda_{\rm cor} =1.67$ was determined by matching the peak stress of the experiment with $3.2$~$\%$ of corrosion.

\begin{figure}
\begin{center}
\includegraphics[width=12cm]{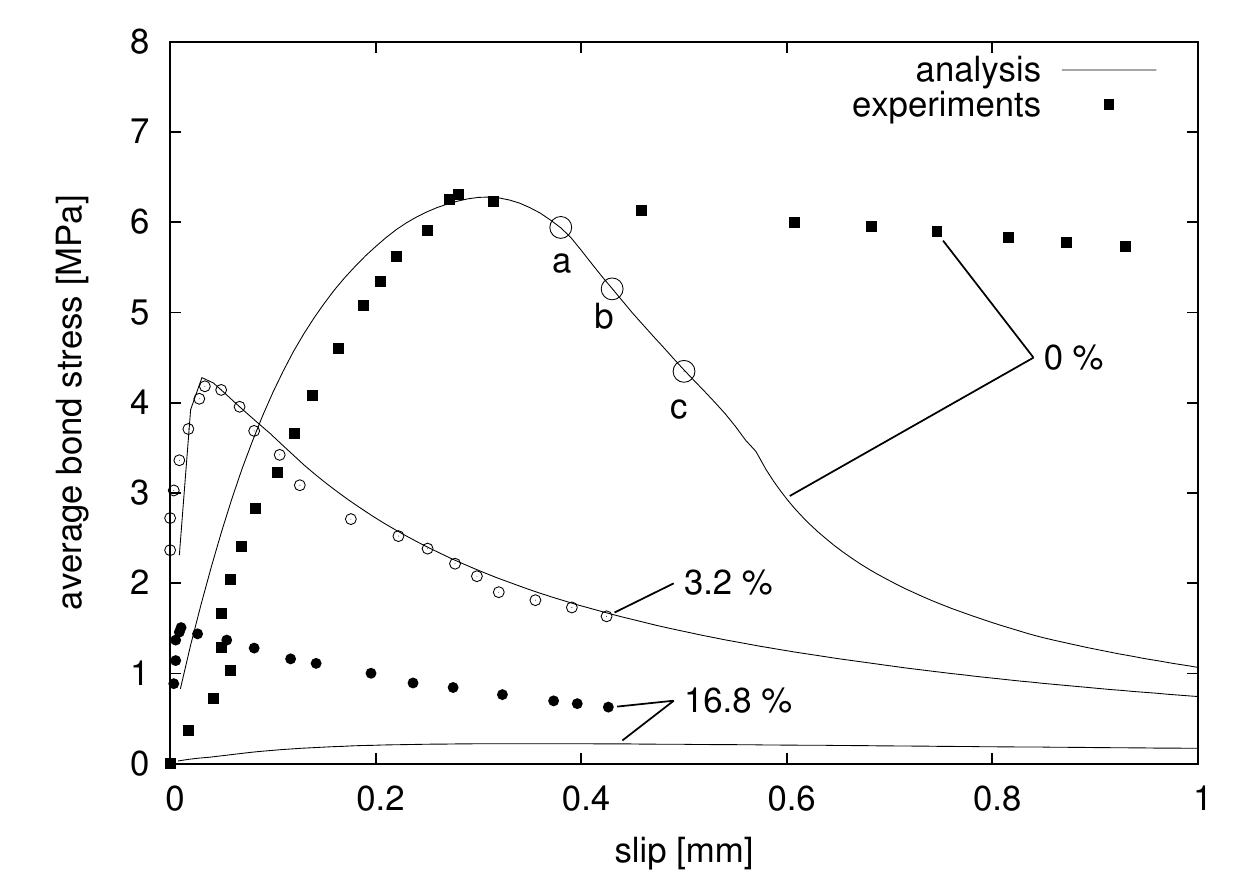}
\end{center}
\caption{Comparison of predicted average bond stress-slip curves and experimental data reported by \protect \cite{LeeNogTom02} for three corrosion percentages $\rho = 0$, $3.2$ and $16.8$~\%.}
\label{fig:ld}
\end{figure}

The results of the analyses are compared to the experimental results in the form of average bond stress-slip curves shown in Fig.~\ref{fig:ld}. Here, the average bond stress was determined as $\tau = F/(\pi \diameter \ell)$, where $\ell = 6 \diameter$ is the embedded length (Fig.~\ref{fig:geometry}a).

\begin{figure}[ht]
\begin{center}
\begin{tabular}{ccc}
\includegraphics[width=5cm]{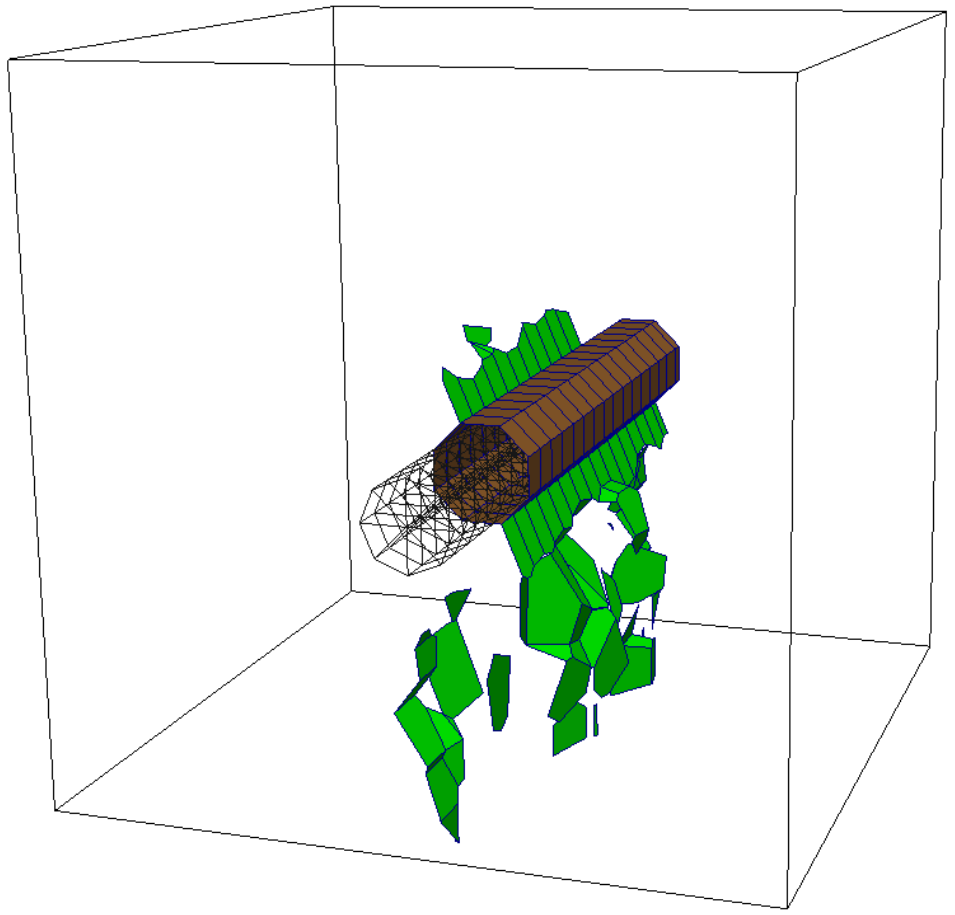} & \includegraphics[width=5cm]{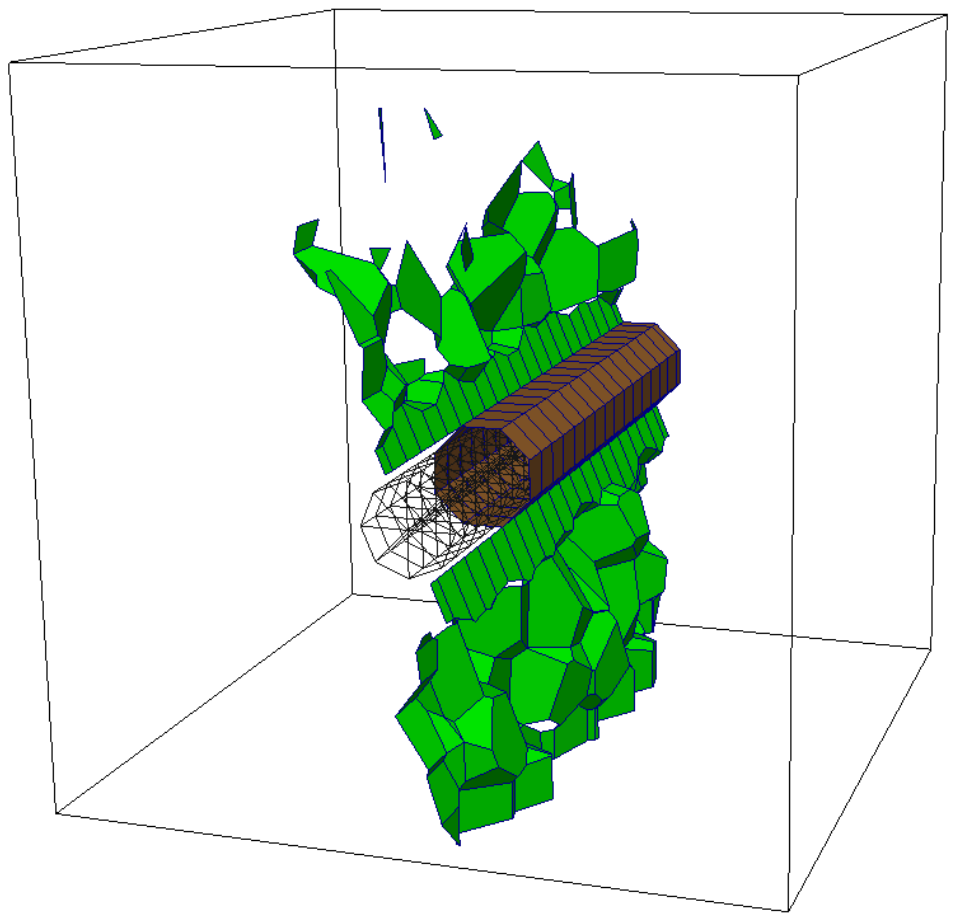} &  \includegraphics[width=5cm]{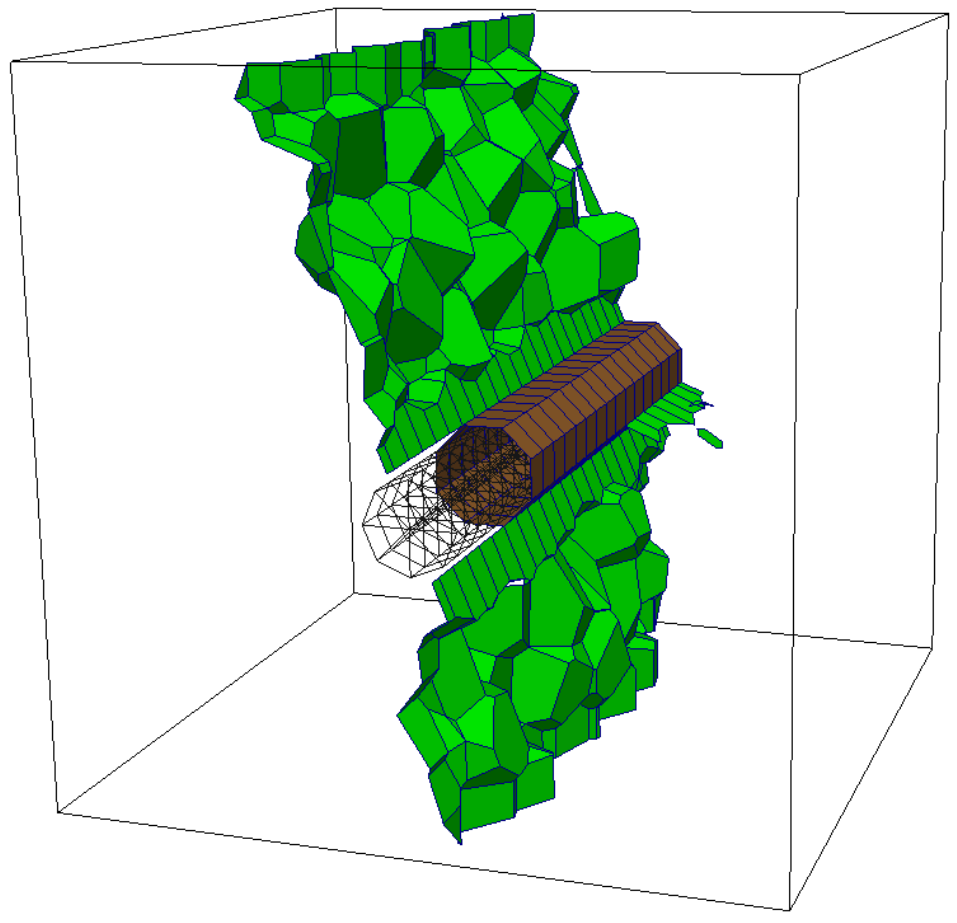}\\
(a) & (b) & (c)\\
\end{tabular}
\end{center}
\caption{Crack patterns for the pullout analysis for the corrosion-free case at (a) $0.38$, (b) $0.42$ and (c) $0.5$~mm of slip.}
\label{fig:crack1}
\end{figure}

\begin{figure}[ht]
\begin{center}
\begin{tabular}{ccc}
 \includegraphics[width=5cm]{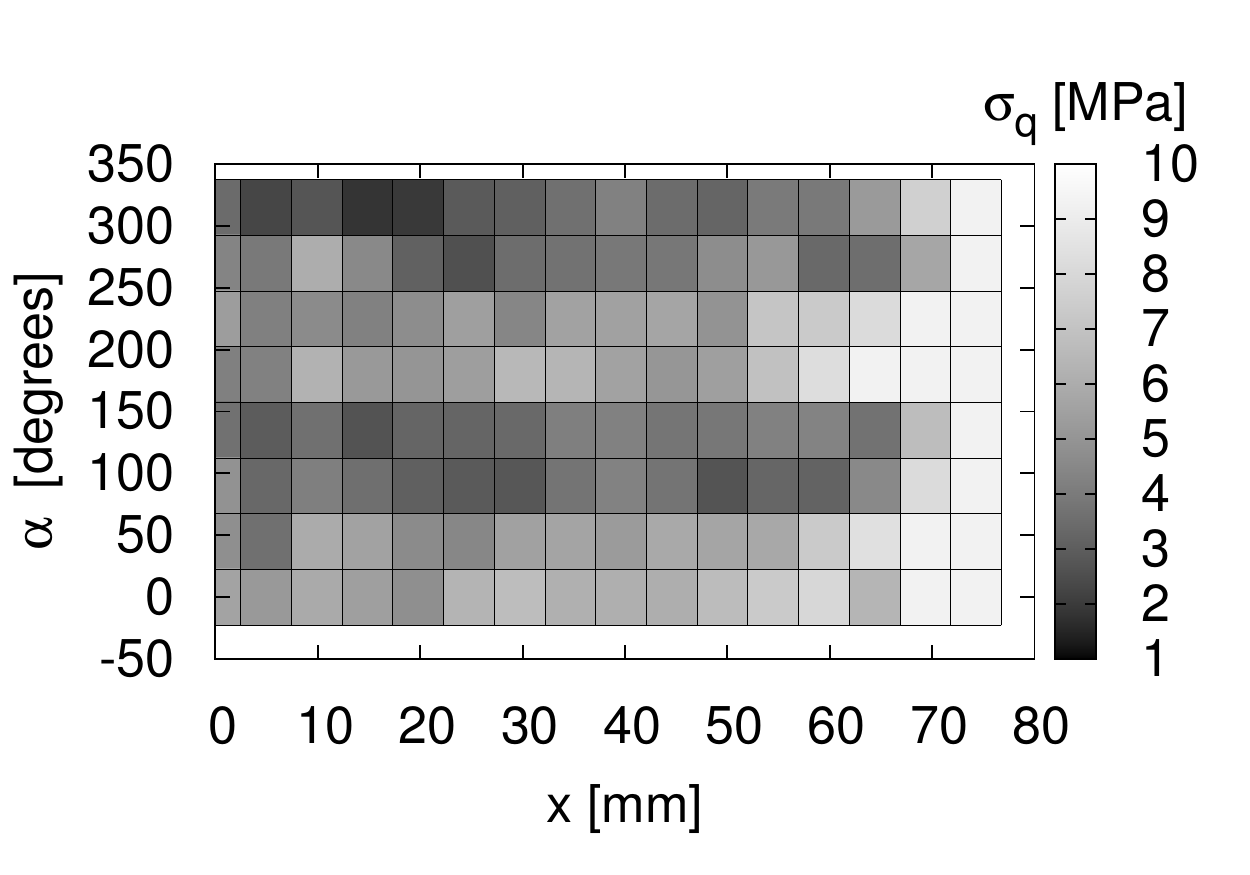} & \includegraphics[width=5cm]{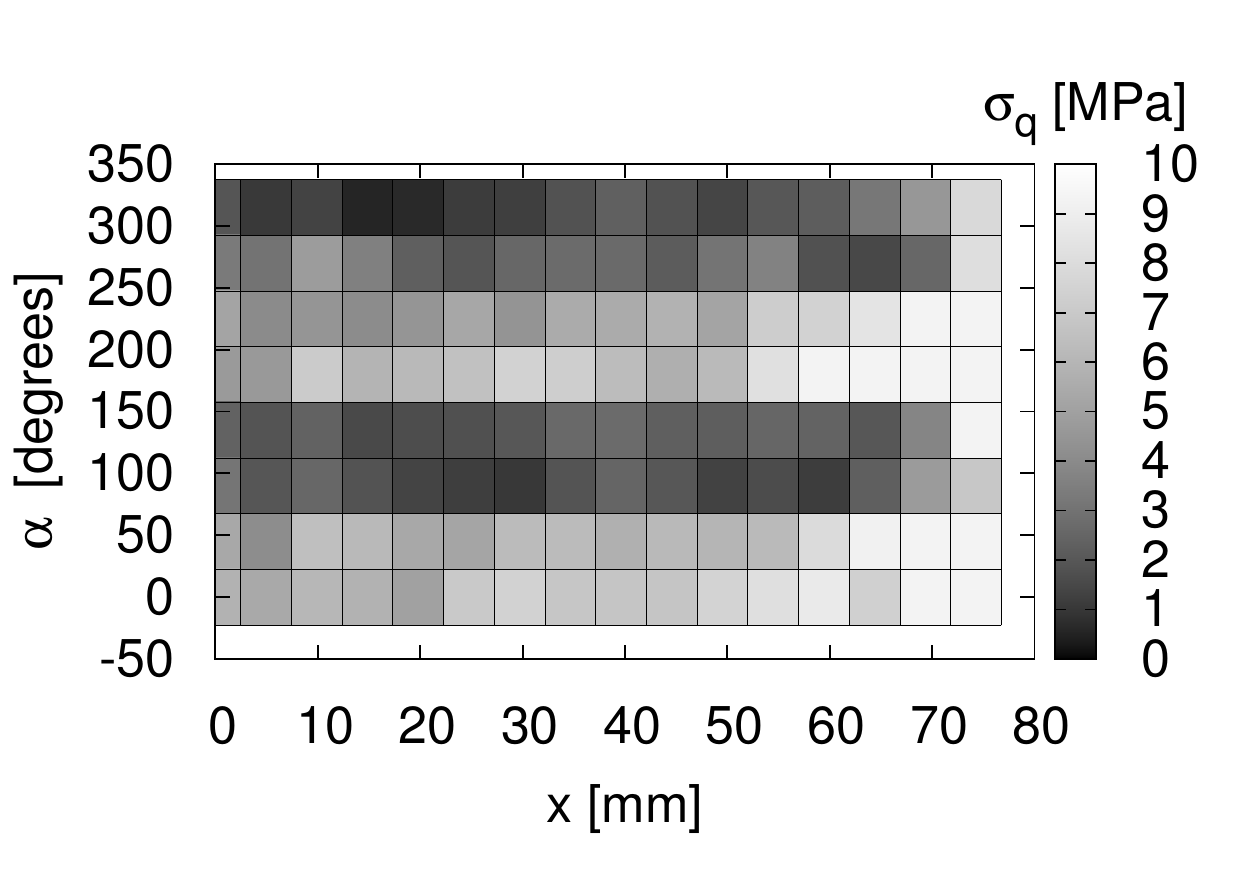} & \includegraphics[width=5cm]{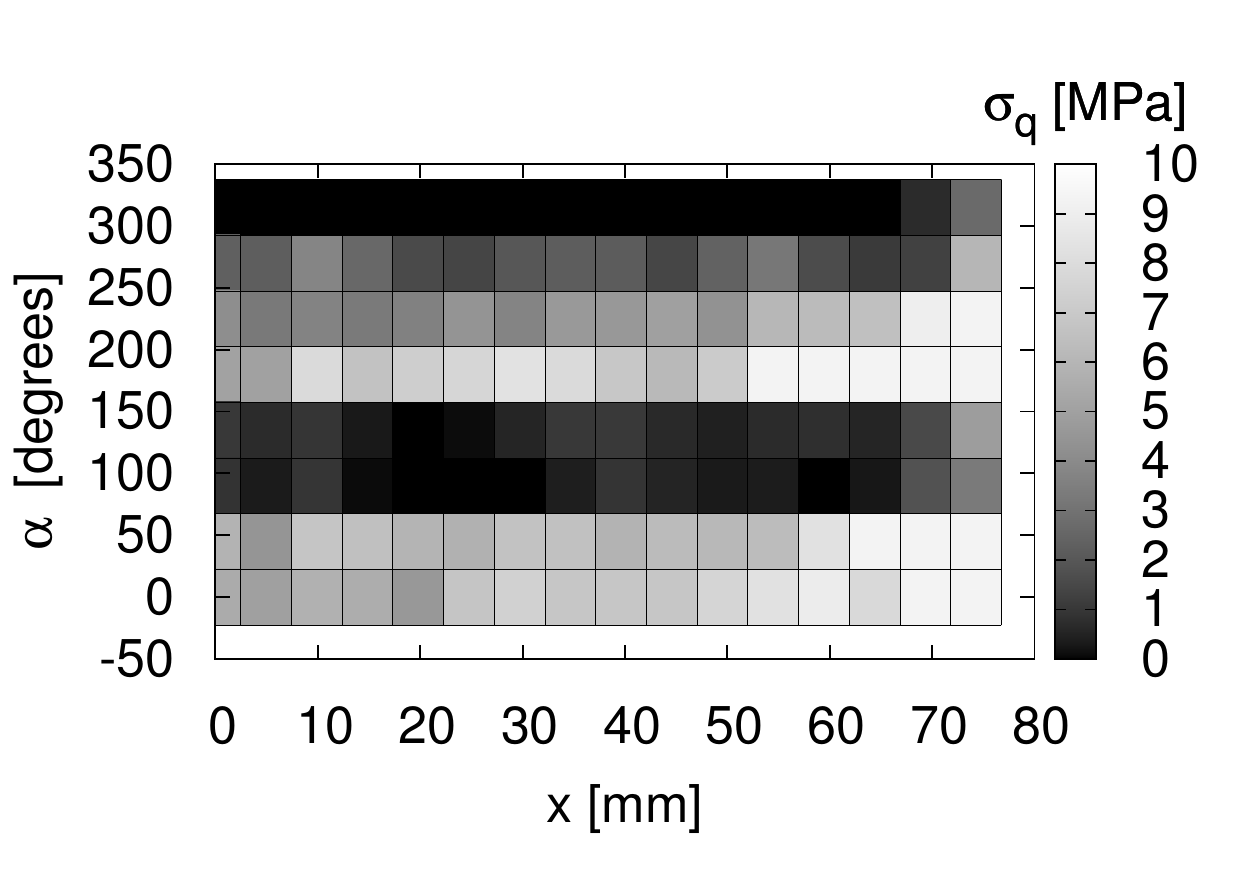}\\
(a) & (b) & (c)
\end{tabular}
\end{center}
\caption{Shear stress distribution along the interface for the three stages marked in Fig.~\ref{fig:ld}.}
\label{fig:shearStress}
\end{figure}

The pre-peak regime of the load-slip curves obtained in the analyses is in very good agreement with the experiments for $0$ and $3.2$~\%, which is expected as these experimental results were used to determine the input parameters.
However, the peak of the analysis for $16.8$~\% strongly underestimates the experimental result. 
Also, for all three corrosion percentages the post-peak response of the analyses is more brittle than the one observed in the experiments.
The main failure mechanism observed in the analysis is the occurrence of splitting cracks which reduce the capacity of the specimen to resist transverse tensile stresses generated by the slip between reinforcement and concrete.
This agrees with the observations made in the experimental study reported in \cite{LeeNogTom02}.
For the analyses without corrosion, the crack patterns for slips of 0.38, 0.42 and 0.5~mm (marked in Fig.~\ref{fig:ld}) are shown in Fig.~\ref{fig:crack1}. 
Crack patterns are visualised as those middle cross-sections of lattice elements, in which the crack opening increases and exceeds 50~$\mu$m at this stage of the analyses. 
These cracks are called active. 
Just after the peak of the average bond stress-slip curve, the concrete cover cracks at its thinnest section (Fig.~\ref{fig:crack1}a).
With further slip, additional cracks initiate from the reinforcement and propagate through the concrete as shown in Fig.~\ref{fig:crack1}b and c.
The results of this pull-out test are further investigated by studying the distribution of the shear stress $\sigma_{\rm q}$ along the interface.
In Fig.~\ref{fig:shearStress}a-c, the distributions for bond slips of $0.38$, $0.42$ and $0.5$~mm are presented. Here, the coordinate $x$ indicates the position along the reinforcement bar and $\alpha$ is the angle in the polar coordinate system defining the position at the interface between reinforcement bar and the concrete in the y-z plane (see Fig.~\ref{fig:geometry}b).
It can be seen that, as the cracks propagate through the concrete, the shear stress becomes increasingly localized in the ``crack-free'' zones (at $\alpha=0^{\circ}$ and $180^{\circ}$).

\begin{figure}[ht]
\begin{center}
\begin{tabular}{cc}
\includegraphics[width=5cm]{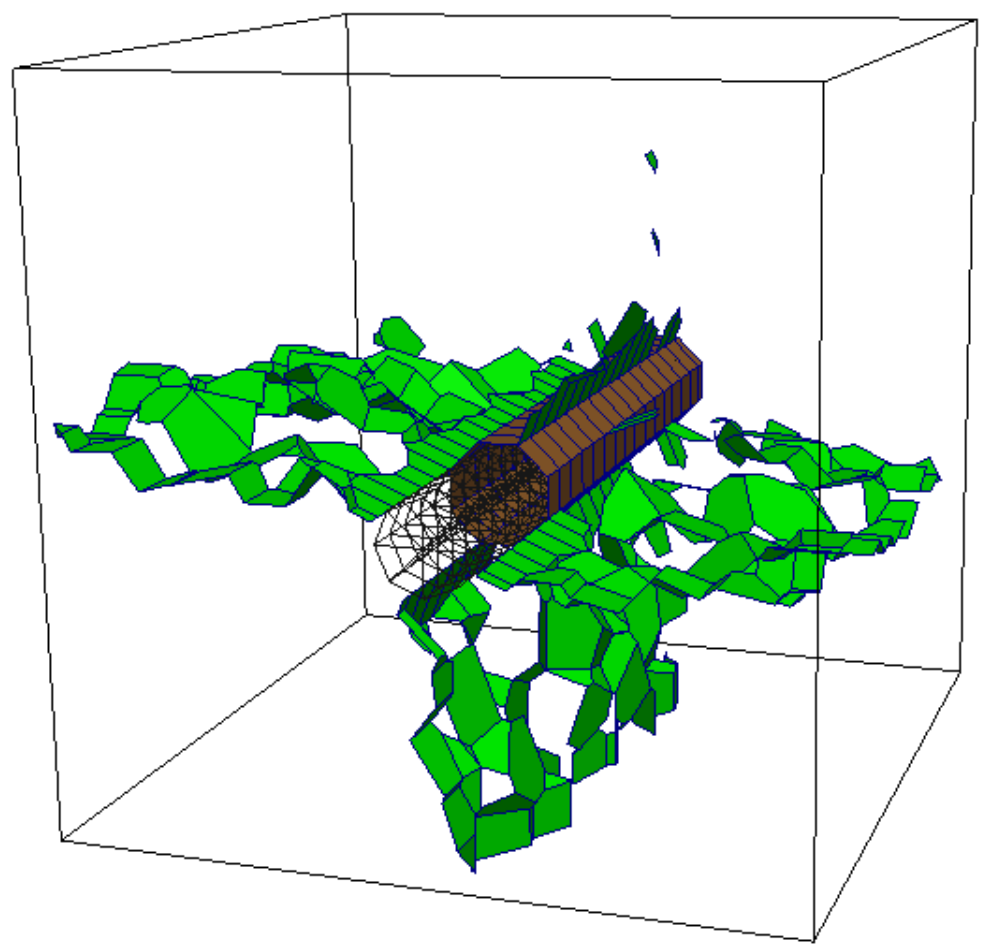} & \includegraphics[width=5cm]{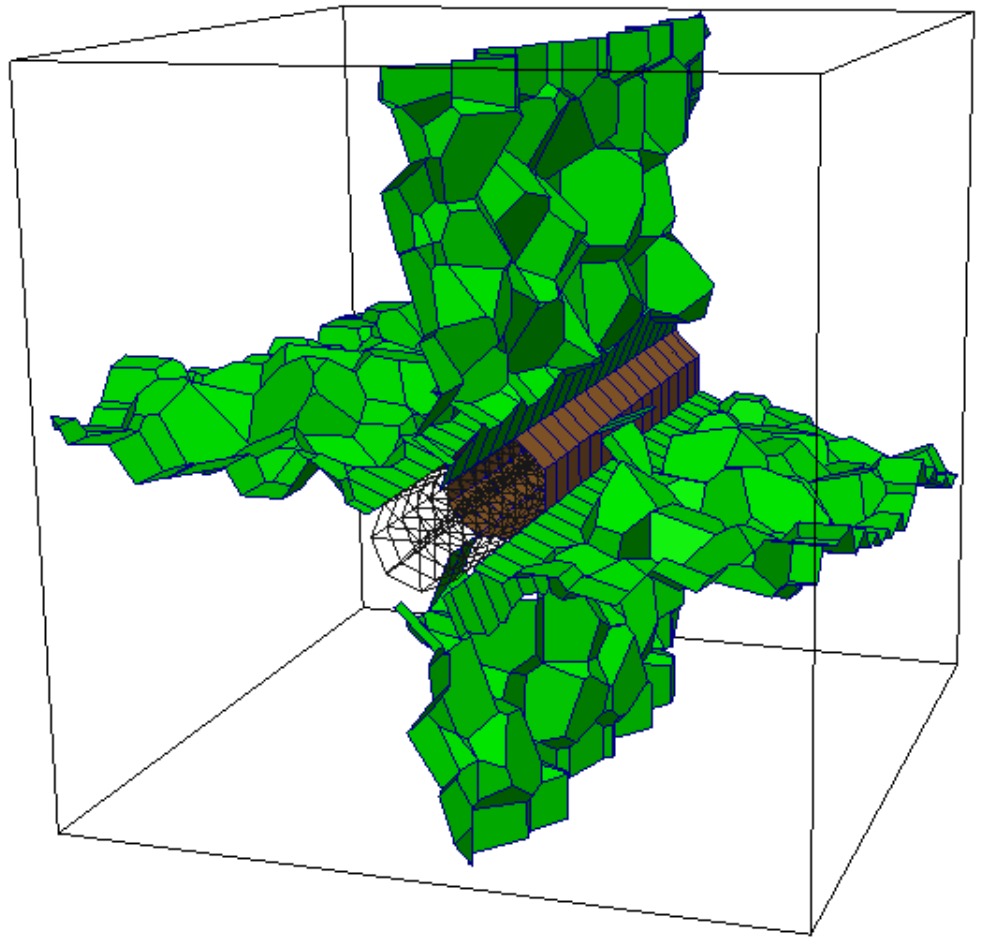}\\
(a) & (b)
\end{tabular}
\end{center}
\caption{Crack patterns for the analyses with (a) $3.2$~$\%$ and (b) $16.8$~$\%$ corrosion before pullout.}
\label{fig:crack2}
\end{figure}

For the corroded bars, the crack patterns are shown in Fig.~\ref{fig:crack2}, at the end of the corrosion process before pull out. Again, only active cracks are shown.
For both these cases, cracking of the concrete cover occurs before pullout, which corresponds to the experimental observations reported in the literature \cite{LeeNogTom02}. 

\begin{figure}
\begin{center}
\includegraphics[width=12cm]{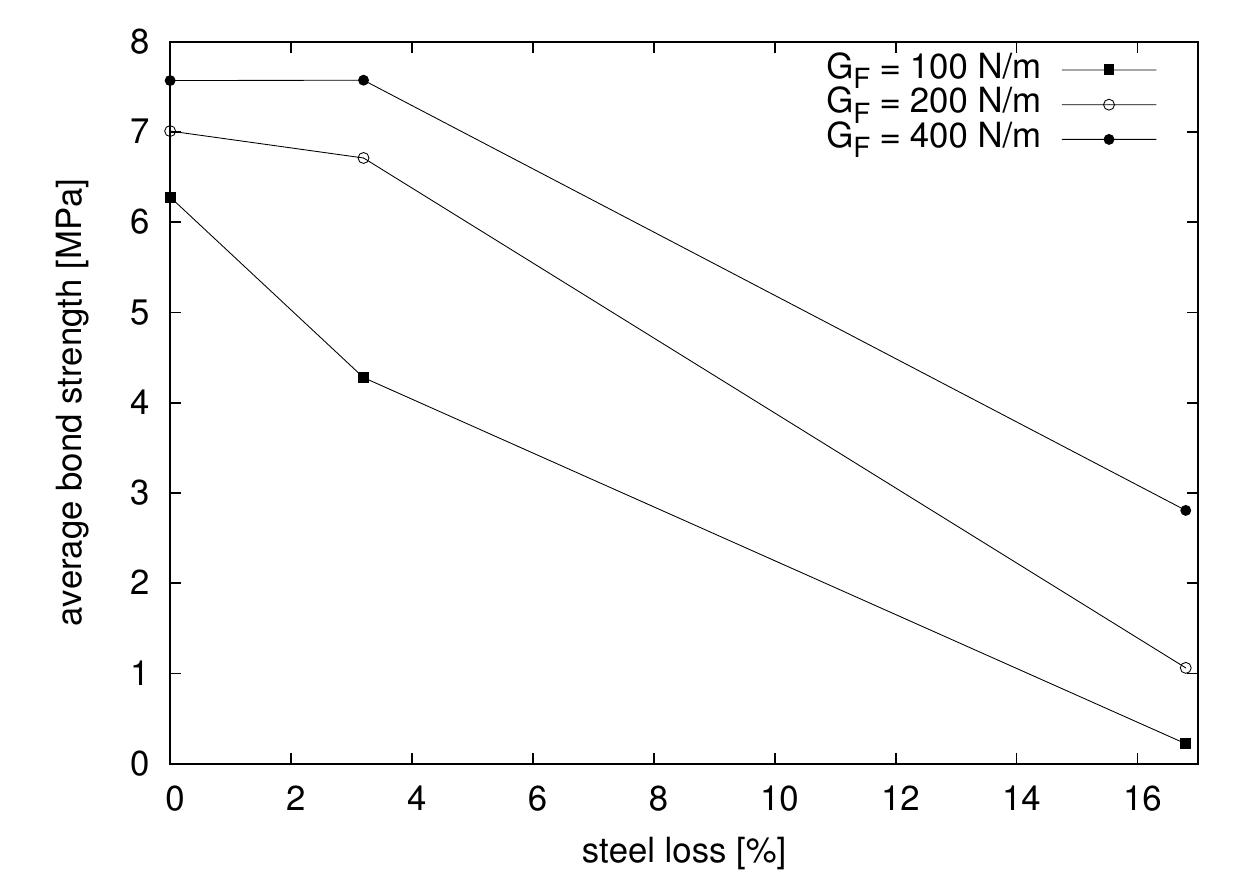}
\end{center}
\caption{Influence of fracture energy on the nominal bond strength.}
\label{fig:fractureEnergy}
\end{figure}

Since the failure mode is dominated by the occurrence of splitting cracks, it is expected that a change in the capacity of the concrete to carry tensile stresses will strongly influence the observed bond behaviour. Therefore, a parametric study was carried out for all three cases, assuming three different fracture energies of the concrete lattice elements, namely $G_{\rm F} = 100$, $200$ and $400$~N/m.
The peak bond stresses obtained from these nine analyses are presented in Fig.~\ref{fig:fractureEnergy}.
It can be seen that the fracture energy strongly influences the bond strength, particularly for highly corroded bars. For instance, an increase of the fracture energy from $100$ to $200$~N/m results in an 4.75~fold increase in nominal bond strength for a corrosion percentage of $16.8$~\%.  
This sensitivity of the results with respect to capacity of the concrete to transmit transverse tensile stresses may well explain the observed discrepancy between the numerical and experimental results, shown in Fig.~\ref{fig:ld}, for the specimen with $\rho = 16.8$~\%. 
For instance, in the analysis, the boundary at the face of the specimen, at which the force was applied, was assumed to be supported only in the direction of the applied force. 
However, it is likely that in the experiments some friction between the loading plate and the concrete specimen was present, which would have provided lateral resistance.
Another simplification is the use of a constant expansion parameter $\lambda_{\rm{cor}}=1.67$ for varying steel loss. 
Experimental results reported in \cite{WonZhaKar10} show that rust products penetrate into the cracked concrete. 
This penetration is expected to be more significant in highly corroded bars.
Furthermore, for highly corroded bars, compaction of the rust product as reported in \cite{OugBerFraFoc06} might play a role as well.

\section{Conclusions}
A lattice approach is used to describe the mechanical interaction of a corroding reinforcement bar, the surrounding concrete and the interface between steel reinforcement and concrete. The cross-section of the ribbed reinforcement bar is taken to be circular, assuming that the interaction of the ribs of the deformed reinforcement bar and the surrounding concrete is included in a cap-plasticity interface model. 
This lattice approach is capable of representing many of the important characteristics of corrosion-induced cracking and its influence on bond.
The idealisation of the corrosion expansion as an Eigenstrain allows for the modelling of corrosion-induced cracking.
The frictional bond law can model the decrease of the bond strength if the concrete is pre-cracked.
Good agreement with experimental results in the pre-peak regime of the bond stress-slip curves was obtained but more studies are required to investigate the post-peak response of the bond stress-slip curves.
It should be emphasised that the parameter controlling the amount of expansive Eigenstrain is not a material parameter of the rust, but a model parameter which takes into account the effect of many micro and meso-structure effects, such as compaction of the rust and penetration of rust into cracks. In future studies, the dependence of this parameter on the amount of corrosion will be investigated.

\section{Acknowledgements}
The authors would like to express their gratitude to Dr. Bo\v{r}ek Patz\'{a}k of the Czech Technical University for kind assistance with his finite element package OOFEM \citep{Pat99,PatBit01}.

\bibliographystyle{plainnat}
\bibliography{general}

\clearpage

\end{document}